\documentclass[useAMS,usenatbib,usegraphicx]{mn2e}
\usepackage{aas_macros,amssymb}
\title{Absolutely calibrated radio polarimetry of the inner Galaxy at 2.3~GHz and 4.8~GHz}
\author[X. H. Sun et al.]
  {\parbox{\textwidth}{X. H.~Sun$^1$\thanks{E-mail: xiaohui.sun@sydney.edu.au}, 
   B. M.~Gaensler$^1$, E.~Carretti$^2$, C. R.~Purcell$^1$, L. Staveley-Smith$^{3,\,4}$, G. Bernardi$^{5,\,6,\,7}$ and M. Haverkorn$^8$}\vspace{0.4cm}\\
  \parbox{\textwidth}{
  $^1$Sydney Institute for Astronomy, School of Physics, The University of 
      Sydney, New South Wales 2006, Australia\\
  $^2$CSIRO Astronomy and Space Science, PO Box 276, Parkes, New South Wales 2870, Australia\\
  $^3$International Centre for Radio Astronomy Research, M468, University of Western Australia, 35 Stirling Highway, Crawley, Western Australia 6009, Australia\\
  $^4$ARC Centre of Excellence for All-sky Astrophysics (CAASTRO), M468, University of Western Australia, 35 Stirling Highway, Crawley, Western Australia 6009, Australia\\
  $^5$SKA SA, 3rd Floor, The Park, Park Road, Pinelands, 7405, South Africa\\
  $^6$Department of Physics and Electronics, Rhodes University, PO Box 94, Grahamstown, 6140, South Africa\\
  $^7$Harvard-Smithsonian Center for Astrophysics, Garden Street 60, Cambridge, MA, 02138, USA\\
  $^8$Department of Astrophysics/IMAPP, Radboud University Nijmegen, PO Box 9010, NL-6500 GL Nijmegen, the Netherlands}
}

\date{Received xxx}
\begin{document}
\maketitle
\begin{abstract}

We present high sensitivity and absolutely calibrated images of diffuse radio 
polarisation at a resolution of about $10\arcmin$ covering the range 
$10\degr<l<34\degr$ and $|b|<5\degr$ at 2.3~GHz from the S-band Parkes All Sky 
Survey and at 4.8~GHz from the Sino-German $\lambda$6~cm polarisation 
survey of the Galactic plane. Strong depolarisation near the Galactic plane is 
seen at 2.3~GHz, which correlates with strong H$\alpha$ emission. We ascribe 
the depolarisation to spatial Faraday rotation measure fluctuations of about 
65~rad~m$^{-2}$ on scales smaller than 6--9~pc. We argue that most (about 90\%) 
of the polarised emission seen at 4.8~GHz originates from a distance of 
3--4~kpc in the Scutum arm and that the random magnetic field dominates the 
regular field there. A branch extending from the North Polar Spur towards lower 
latitudes can be identified from the polarisation image at 4.8~GHz but only 
partly from the polarised image at 2.3~GHz, implying the branch is at a 
distance larger than 2--3~kpc. We show that comparison of structure functions 
of complex polarised intensity with those of polarised intensity can indicate 
whether the observed polarised structures are intrinsic or caused by Faraday 
screens. The probability distribution function of gradients from the 
polarisation images at 2.3~GHz indicates the turbulence in the warm ionised 
medium is transonic. 
\end{abstract}
\begin{keywords}
ISM: magnetic fields -- polarisation -- radio continuum: general -- radio 
continuum: ISM
\end{keywords}

\section{introduction}

Polarised radio emission provides information on both the magnetic field at 
locations where the emission is generated and the magnetic field that pervades 
the ionised interstellar medium through which the emission then propagates. 
Therefore radio polarisation observations provide a unique probe of the 
Galactic magnetic field coupled with both thermal and relativistic electrons.

The magneto-ionic medium (MIM) modulates the linearly polarised emission that 
propagates through it via Faraday rotation. The polarisation angle of the 
emission is rotated by an amount proportional to the wavelength squared, with 
the coefficient defined as rotation measure (RM). RM is calculated as the 
integral of the magnetic field strength parallel to the line of sight weighted 
by thermal electron density and the integral is assessed from the source to the 
observer. Diffuse polarised emission at different locations along the line of 
sight or across the observation beam can experience different rotations, and 
adding up all the emission can thus cancel partly or completely the 
polarisation \citep[``depolarisation"; e.g.][]{sbs+98}. 

The degree of depolarisation clearly depends on observational frequency. The 
higher the frequency, the weaker the depolarisation and the deeper we can thus 
penetrate into the Galactic interstellar medium. This implies that 
multi-frequency polarisation observations can let us extract a series of slices 
from the MIM and thus gain a 3D view of the MIM. 

Depolarisation also depends on direction. In the Galactic plane, particularly 
towards the inner Galaxy, RMs and their spatial variations are both very large 
\citep[e.g.][]{bhg+07} because lines of sight cross many spiral arms. 
Therefore high frequency observations are required to detect diffuse polarised 
emission coming from large distances near the Galactic plane. In contrast, 
towards high Galactic latitudes, RMs are much smaller \citep{tss09} and 
diffuse polarised emission can be observed at low frequencies 
\citep[e.g.][at 189~MHz]{bgm+13}.  

Currently there are many radio polarisation surveys available covering a broad 
frequency range. Examples of Galactic plane polarisation surveys are the 
Sino-German $\lambda$6~cm polarisation survey \citep[e.g.][]{srh+11}, the 
Effelsberg 2.7~GHz survey~\citep{drrf99}, the Parkes 2.4~GHz survey 
\citep{dhjs97}, the Canadian Galactic Plane Survey at 1.4~GHz 
\citep[CGPS, e.g.][]{lrr+10}, and the Southern Galactic Plane Survey at 1.4~GHz 
\citep[SGPS, e.g.][]{hgm+06}. Examples of all-sky radio polarisation surveys 
including the Galactic plane are the WMAP all-sky surveys \citep{hwh+09}, the 
S-band Polarisation All Sky Survey \citep[S-PASS,][]{car10}, and the 1.4~GHz 
all-sky survey \citep{wlrw06,trr08}. Ongoing surveys such as the Galactic 
Magneto-Ionic Medium Survey \citep[GMIMS,][]{wlh+10} and the Galactic Arecibo 
L-band Feeds Array Continuum Transit Survey \citep[GALFACTS,][]{ts10} using 
multi-channel polarimetry and RM synthesis \citep{bd05}, will be able to reveal 
polarised emission at a wide range RMs, and therefore will further advance our 
understanding of the Galactic interstellar medium.

All the Galactic plane surveys share a common feature -- most of the structures 
seen in polarisation, such as ``voids", ``canals" and ``patches", do not have 
corresponding structures in total intensity, irrespective of frequency, 
resolution and telescope (whether single dish or synthesis array). Broadly, 
there are two possible mechanisms for these structures. The first possibility 
is that the random magnetic fields dominate over the regular fields in the 
emitting region and produce the observed polarised 
intensities~\citep[e.g.][]{srh+11}. The random fields can be either anisotropic 
\citep[e.g.][]{fbs+11,jbl+11,jf12} or isotropic but with a non-zero energy 
spectrum~\citep{sr09}. The second possibility is that the background polarised 
emission is smooth in spatial distribution, but foreground Faraday screens 
impose Faraday rotation and hence produce the structures in polarisation. The 
Faraday screens could be either diffuse warm ionised medium with jumps or cusps 
in electron density caused by MHD turbulence \citep{ghb+11} or discrete objects 
such as H{\scriptsize II} regions and others whose properties still remain 
unclear \citep{gdm+01,srh+11}. The first scenario is intrinsic to the emitting 
regions and the second is extrinsic. The two mechanisms are best diagnosed at 
high and low frequencies, respectively. This reinforces the importance of 
having a wide frequency coverage to simultaneously investigate both mechanisms. 

In this paper, we focus on the region $10\degr<l<34\degr$ and $|b|<5\degr$, 
which is covered by both S-PASS \citep{car10} and the Urumqi survey 
\citep{srh+11} providing absolutely calibrated, high resolution and sensitivity 
observations at 2.3~GHz and 4.8~GHz. This is the only part of the Galactic 
plane where such kinds of dataset are currently available, allowing us to 
investigate different layers of the MIM. These two frequencies are ideal as the 
high frequency ensures intrinsic polarised emission can be detected while the 
low frequency results in depolarisation, so that different layers can be 
separated.

The paper is organised as follows. We show the observational data and describe 
the polarisation images in Sect~2. We present the discussions on diffuse 
polarised emission, the branch extending from the North Polar Spur (NPS) and 
statistical studies of the polarisation images in Sect.~3. We give our 
conclusions in Sect.~4. Throughout the paper, all coordinates are Galactic and 
the intensities are presented as main beam brightness temperatures 
\citep[see e.g.][]{wrh12}. 

\begin{figure*}
\centering
\includegraphics[angle=-90,width=0.9\textwidth]{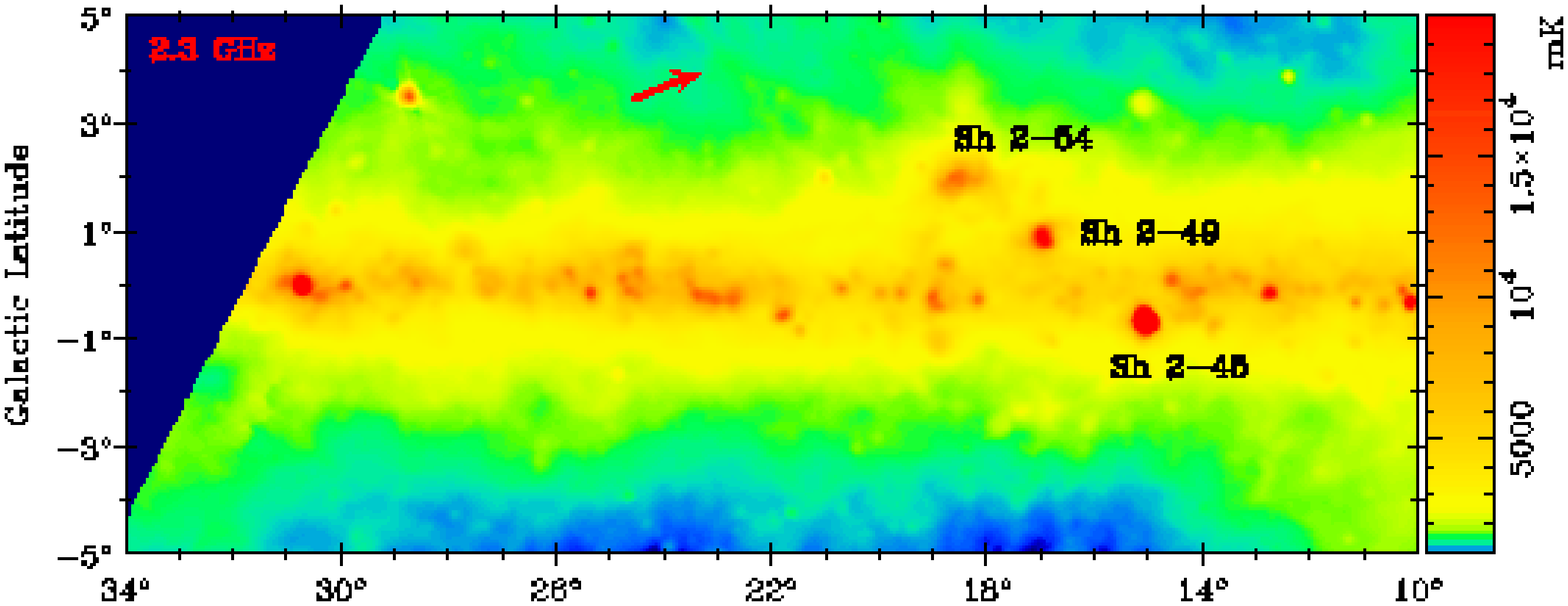}\\[2mm]
\includegraphics[angle=-90,width=0.9\textwidth]{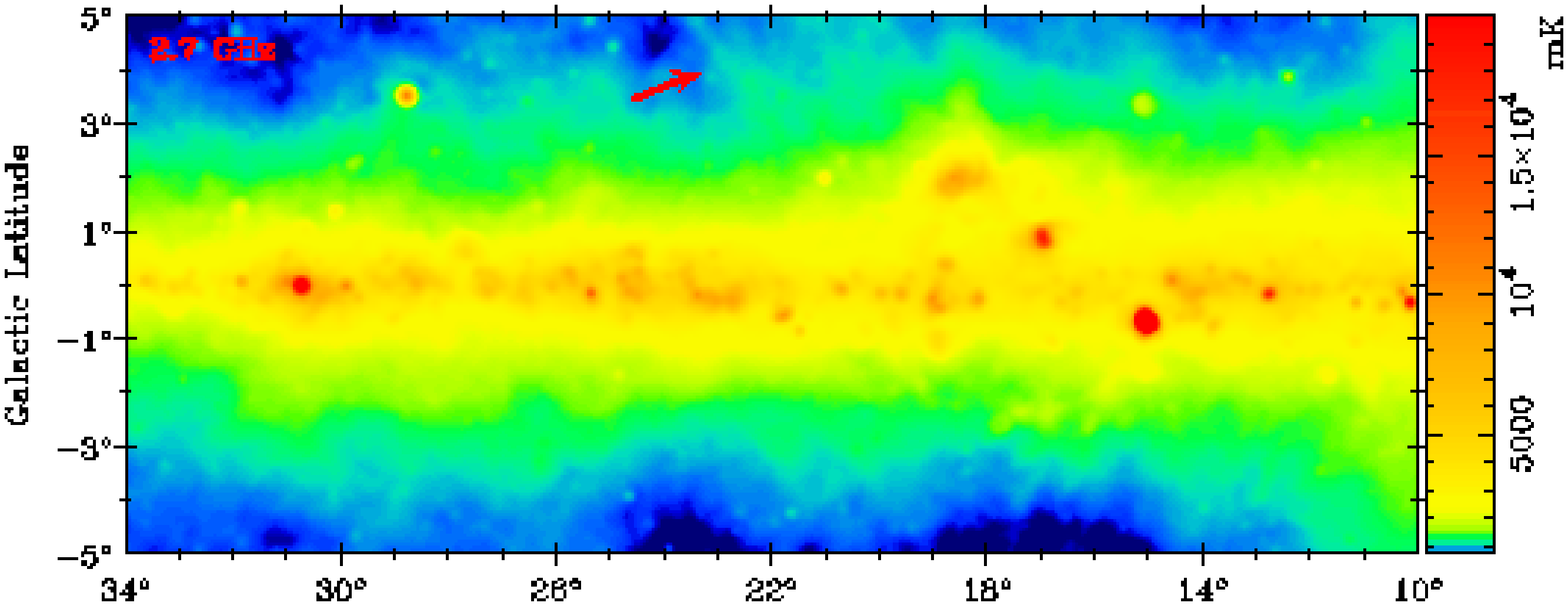}\\[2mm]
\includegraphics[angle=-90,width=0.9\textwidth]{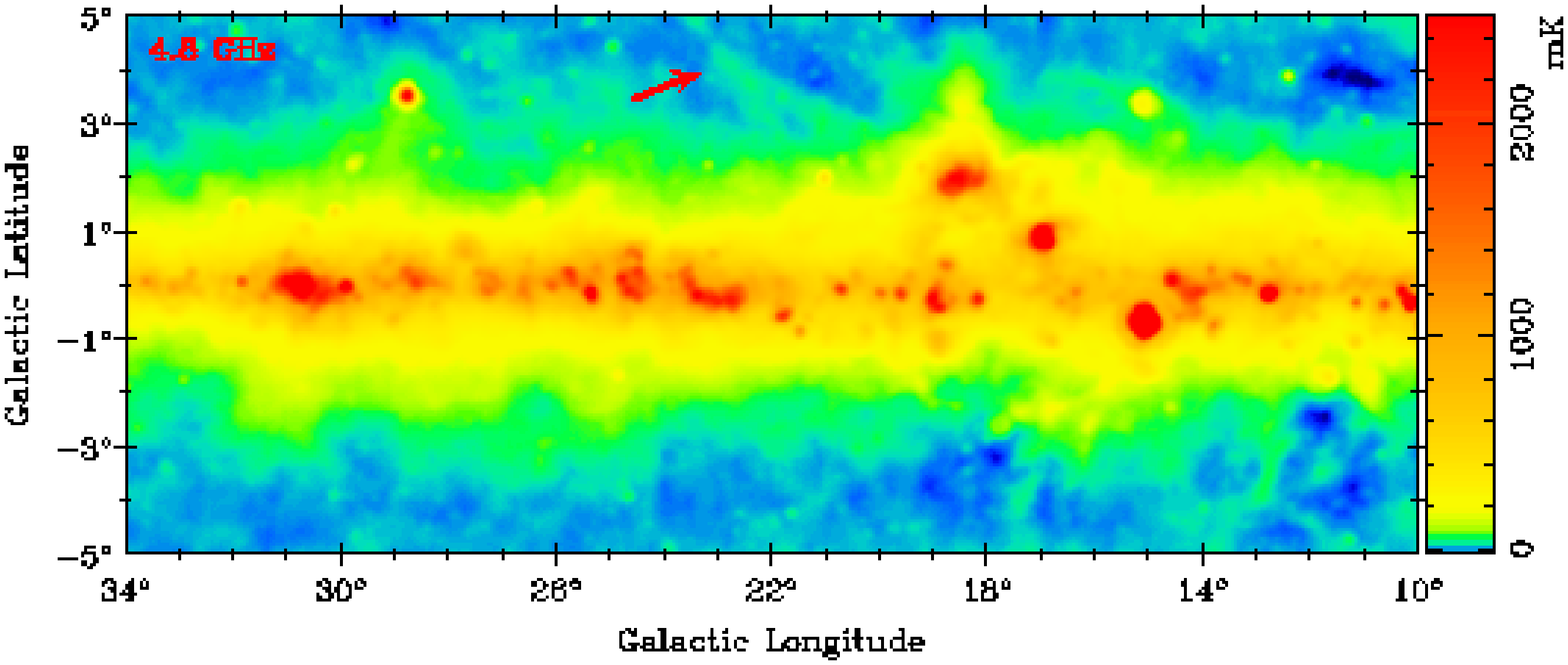}
\caption{Total intensity images. From top to bottom are: the Parkes 2.3~GHz 
data, the Effelsberg 2.7~GHz data and the Urumqi 4.8~GHz data. All maps have 
been smoothed to a common resolution of $10\farcm75$. Three bright 
H{\scriptsize II} regions are marked. The arrows point at the branch extending 
from the North Polar Spur.}
\label{toti}
\end{figure*}

\begin{figure*}
\centering
\includegraphics[angle=-90,width=0.9\textwidth]{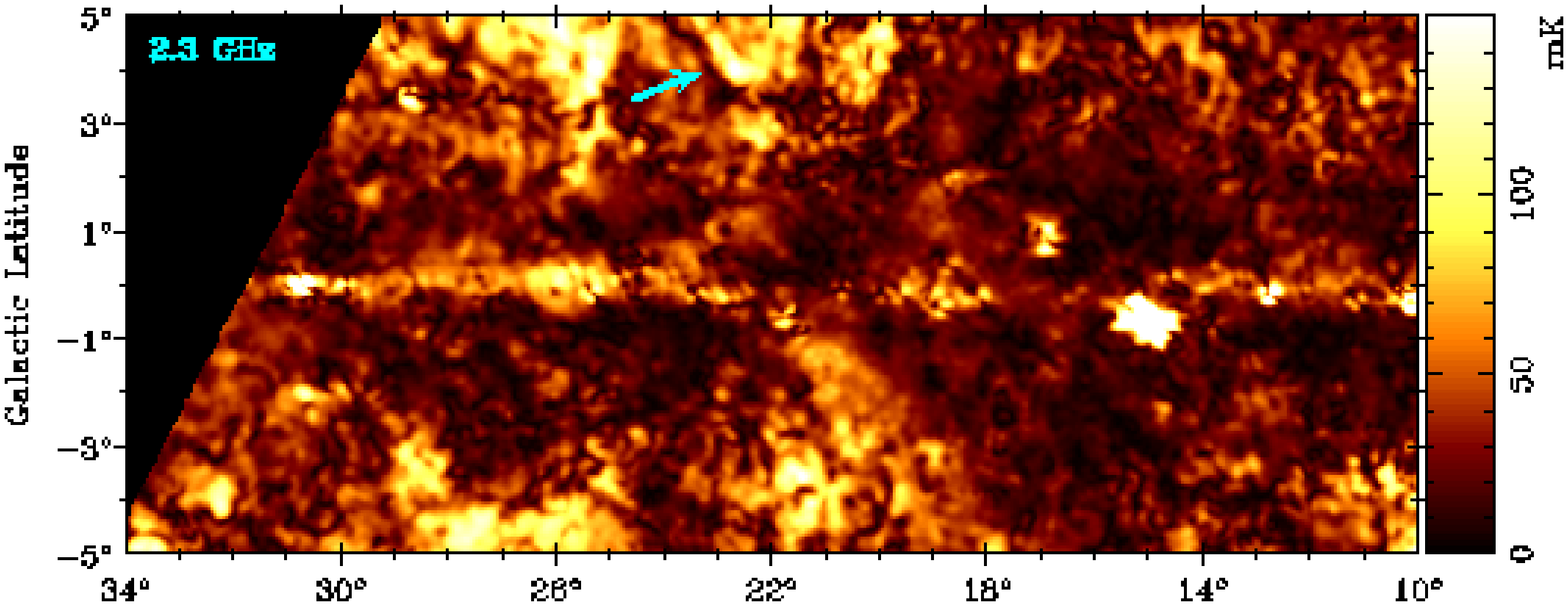}\\[2mm]
\includegraphics[angle=-90,width=0.9\textwidth]{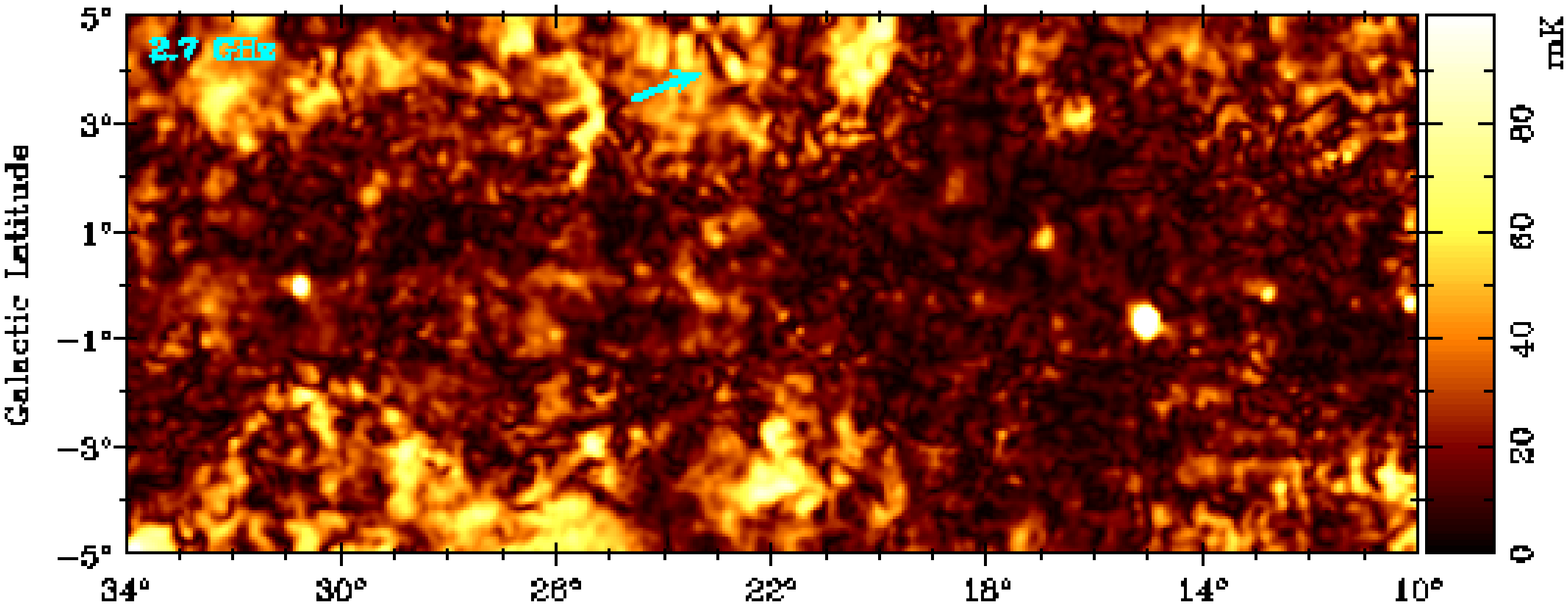}\\[2mm]
\includegraphics[angle=-90,width=0.9\textwidth]{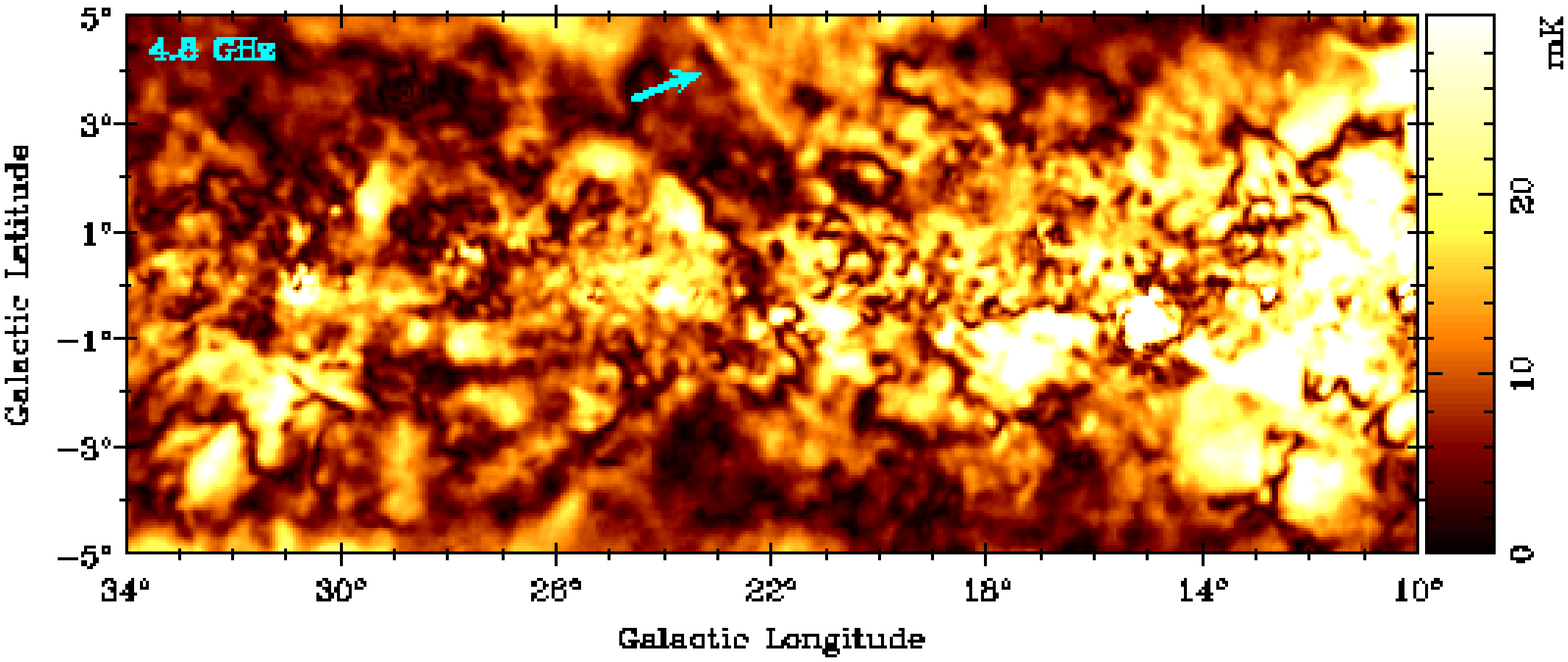}
\caption{The same as Fig.~\ref{toti} but for images of linearly polarised 
intensity. The arrows point at the branch extending from the North Polar Spur.}
\label{pi}
\end{figure*}

\section{Observational Data}

We show both total intensity and polarisation data in this section, but most of 
the discussion will be focused on polarisation. We mainly use 2.3~GHz data 
taken from the S-PASS \citep{car10} observed with the Parkes 64-m telescope and 
4.8~GHz data taken from the Sino-German $\lambda$6~cm polarisation survey of 
the Galactic plane \citep{srh+11} observed with the Urumqi 25-m telescope. 
The polarisation data at both 2.3~GHz and 4.8~GHz have been absolutely 
calibrated to cover structures of all angular scales down to the resolution. We 
also investigate 2.7~GHz data taken from the Galactic plane survey 
\citep{drrf99} with the Effelsberg 100-m telescope.

\subsection{The Parkes 2.3~GHz data}

A brief introduction to S-PASS was given by \citet{car10} and a detailed 
description as well as the data release will be presented in a forthcoming 
paper (Carretti et al. in prep.). The observations were conducted in fast long 
azimuthal scans and were absolutely calibrated by modelling variations of $U$ 
and $Q$ versus parallactic angle \citep{car10}. The original resolution was 
$8\farcm9$. In this paper $I$, $Q$ and $U$ data have all been slightly smoothed 
to $10\farcm75$, and the resulting rms sensitivity in polarised intensity is 
about 0.3~mK per beam-sized pixel. 

\subsection{The Urumqi 4.8~GHz data}

The Sino-German $\lambda$6~cm polarisation survey of the Galactic 
plane\footnote{http://zmtt.bao.ac.cn/6cm} was summarised by \citet{hrs+12} and 
the results for the longitude range $10\degr<l<60\degr$ were presented by 
\citet{srh+11}. The observations consist of many longitude scans of typical 
size $2\degr\times10\degr$ and latitude scans of typical size 
$7\degr\times2\fdg6$. The final maps after combining these orthogonal scans 
miss structures larger than about $10\degr$. Absolute calibration was derived 
by tying the levels of large-scale structures in $U$ and $Q$ to the WMAP K-band 
(22.8~GHz) data, through a process described in detail by \citet{srh+11}. In 
this paper, we redo the absolute calibration using the recently released 
nine-year WMAP data \citep{blw+13}. The original resolution is $9\farcm5$. We 
have slightly smoothed the data to the same resolution as the Parkes 2.3~GHz 
data. The sensitivity in polarisation is about 1~mK. 

\subsection{The Effelsberg 2.7~GHz data}

The Effelsberg 2.7~GHz survey \citep{rfrr90,drrf99} extended the earlier survey 
by \citet{jfr87} from $|b|=1\fdg5$ to $|b|=5\degr$. The data are public at the 
MPIfR Survey Sampler\footnote{http://www.mpifr-bonn.mpg.de/survey.html}. The 
observations were composed of many sections of $3\degr\times3\degr$. The 
intensity levels of the final maps were set to satisfy the average of $U$ and 
$Q$ to be zero. No absolute calibration has been made to the data. The original 
resolution was $4\farcm3$. The data have also been smoothed to a resolution of 
$10\farcm75$ and the sensitivity in polarisation is about 6~mK.

\subsection{Main features}

The total intensity maps for the Parkes 2.3~GHz, the Effelsberg 2.7~GHz and the 
Urumqi 4.8~GHz data are shown in Fig.~\ref{toti} from top to bottom. With the 
exception that more diffuse emission at high latitudes can be seen at 2.3~GHz 
than at 2.7~GHz and 4.8~GHz, the three maps show nearly the same features. 
Discrete sources including extragalactic sources, H{\scriptsize II} regions and 
supernova remnants are embedded in a bright ridge extending from the middle 
Galactic plane to high latitudes.

The images of linearly polarised emission ($p=\sqrt{Q^2+U^2}$) from the three 
surveys are displayed in Fig.~\ref{pi}. The thin layer of polarised intensity 
at 2.3~GHz around the Galactic plane is caused by instrumental polarisation 
leakage. The two bright polarisation blobs appearing in all the images at 
$(15\fdg0,\,-0\fdg7)$ and $(16\fdg9,\,+0\fdg8)$ are residuals of off-axis 
instrumental polarisation leakage from the bright H{\scriptsize II} 
regions Sh 2-45 and Sh 2-49, which is less than 1.5\% at 2.3~GHz and less 
than 1\% at 4.8~GHz. The on-axis polarisation leakage is about 0.4\% at 
2.3~GHz \citep[e.g.][]{chm+10} and 0.7\% at 4.8~GHz. A detailed description of 
the instrumental polarisation for S-PASS will be presented in a forthcoming 
paper (Carretti et al. in prep.). For diffuse emission, the instrumental 
polarisation leakage is generally not important. 

We can see little resemblance between the total intensity and polarisation 
images at each individual frequency, as has already been noted in many previous 
surveys. In contrast with total intensity, polarisation images show totally 
different morphologies at different frequencies.

Polarisation patches as well as low polarisation voids and canals prevail in 
the Urumqi 4.8~GHz data. In contrast, in the Parkes 2.3~GHz and Effelsberg 
2.7~GHz data, there is nearly no polarisation near the Galactic plane. 
Polarised emission towards a branch which is the low latitude extension of the 
North Polar Spur (NPS) can be identified in the Urumqi 4.8~GHz data 
from $(23\fdg5,\,+5\degr)$ to $(21\fdg7,\,+2\fdg7)$ (Fig.~\ref{pi}). However, 
in the Parkes 2.3~GHz data, only fragments of the upper part of the polarised 
branch are visible. It is also interesting to note that the Effelsberg 2.7~GHz 
image looks very similar to the Parkes 2.3~GHz image although the former is not 
absolutely calibrated. These points will be discussed in Sect.~\ref{discussion}.
 
\section{Discussion}\label{discussion}

In this section, we try to answer the following questions: What causes the 
polarised structures in the Parkes 2.3~GHz and the Urumqi 4.8~GHz data? Can we 
constrain properties of the Galactic magnetic field from these polarised 
structures? Is the NPS local or much further away? Why do the 2.7~GHz data 
without absolute calibration look similar to the 2.3~GHz data with absolute 
calibration? Are the 2.3~GHz and 4.8~GHz data statistically different? 

\subsection{The origin of polarised structures}\label{polgen}

\subsubsection{Polarised structures in the Urumqi 4.8~GHz data}\label{piuru}

\begin{figure}
\centering
\includegraphics[angle=-90, width=0.45\textwidth]{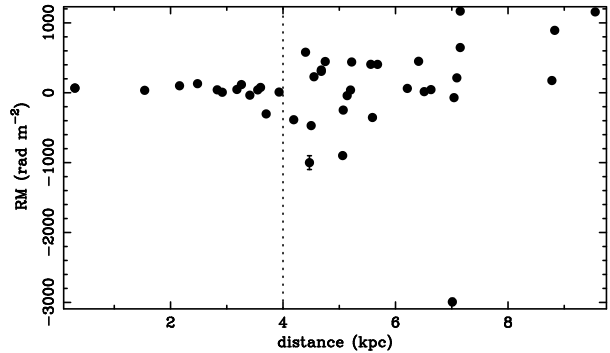}
\caption{Rotation measure versus distance for pulsars in the range 
$10\degr<l<34\degr$ and $|b|<5\degr$. The dashed line indicates a distance of 
4~kpc.}
\label{psrrm}
\end{figure}

\citet{srh+11} argue that the polarised emission in the longitude range 
$10\degr<l<60\degr$ near the Galactic plane at 4.8~GHz originates from a 
distance less than about 4~kpc. We collected data of pulsars falling in the 
range $10\degr<l<34\degr$ and $|b|<5\degr$ from the ATNF pulsar catalogue 
\citep{mhth05}, version 1.45. RMs of these pulsars versus their distances 
calculated using the NE2001 thermal electron density model \citep{cl02} are 
shown in Fig.~\ref{psrrm}. It is clear that pulsar RMs start to exhibit large 
fluctuations at a distance of about 4~kpc. These fluctuations 
($\sigma_{\rm RM}\approx500$~rad~m$^{-2}$) are sufficient to completely destroy 
polarised emission from beyond 4~kpc at 4.8~GHz either through depth or beam 
depolarisation \citep{sbs+98}. The argument is also supported by simulations 
based on 3D-emission models of the Galaxy \citep{srwe08,sr10} showing that 
polarised emission stops increasing beyond a distance of about 4~kpc 
\citep[][their Fig.~4]{srh+11}. We therefore conclude that the polarised 
emission at 4.8~GHz originates from within 4~kpc.  

\begin{figure}
\centering
\includegraphics[width=0.4\textwidth]{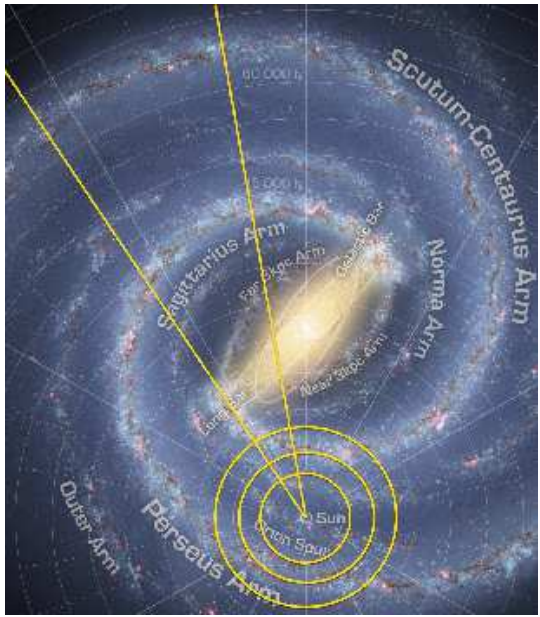}
\caption{Artist's conception of the Milky Way (R. Hurt: NASA/JPL-Caltech/SSC) 
with three yellow circles marking distances of 2~kpc, 3~kpc and 4~kpc. The 
two yellow straight lines indicate longitudes of $10\degr$ and $34\degr$.} 
\label{spiral}
\end{figure}

In Fig.~\ref{spiral}, we indicate the longitude range $10\degr<l<34\degr$ 
covered by both the Parkes 2.3~GHz data and the Urumqi 4.8~GHz data on a sketch 
of the spiral arm pattern of the Milky Way Galaxy which was constructed in 2008 
according to most of the known observational data \citep[e.g. ][]{umh+11}. 
The distance of 4~kpc towards this direction corresponds to the middle or the 
inner edge of the Scutum arm. We will show in Sect.~\ref{dpparkes} that the 
polarised emission at 4.8~GHz actually is confined to the Scutum arm instead of 
being distributed uniformly along lines of sight.

The polarised emission at 4.8~GHz manifests itself as patchy structures. One 
explanation involving Faraday effects is that the polarised emission is 
smoothly distributed with total intensity. However, there might be larger 
foreground RM fluctuations towards some regions arising from irregularities in 
the interstellar medium, and these cause depolarisation in the form of voids 
or canals. The smooth background polarised emission is thus broken into patches.
 According to Fig.~\ref{psrrm}, the RMs for pulsars within 4~kpc have a 
standard deviation of only 44~rad~m$^{-2}$ (the outlier of $-$303~rad~m$^{-2}$ 
was discarded). These RM fluctuations only produce polarisation angle 
fluctuations of about $10\degr$, which is too low to cause depolarisation at 
4.8~GHz. This means that foreground Faraday screens cannot produce the 
polarisation patches observed.

Most of the polarised structures at 4.8~GHz therefore must be intrinsic, and 
either caused by fluctuations in cosmic ray electrons or magnetic fields. With 
a typical lifetime of about $10^7$~yr for cosmic ray electrons emitting at 
4.8~GHz in a magnetic field of several $\mu$G and a diffusion coefficient of 
about $10^{28}$~cm$^2$~s$^{-1}$~\citep{fbs+11}, we derive a diffusion length of 
about 3~kpc. This means cosmic ray electrons are nearly uniformly distributed 
across the spiral arm and inter-arm regions. We therefore ascribe the polarised 
structures to magnetic fields. Following \citet{sbs+98}, we represent total 
intensity ($I$) and polarised intensity ($p$) as
\begin{equation}\label{eq:ipi}
\begin{array}{r@{}c@{}l}
I &\,\,\propto\,\,& \bar{B}_x^2+\bar{B}_y^2+\sigma_x^2+\sigma_y^2\\[2mm]
p &\,\,\propto\,\,& \left[ \left(\bar{B}_x^2-\bar{B}_y^2+\sigma_x^2-\sigma_y^2\right)^2+4\bar{B}_x^2\bar{B}_y^2\right]^{1/2},
\end{array}
\end{equation}
where $\bar{B}_x$ and $\bar{B}_y$ stand for the two orthogonal components of 
average regular magnetic field perpendicular to the line of sight, and 
$\sigma_x$ and $\sigma_y$ stand for the strength of random magnetic fields. 
Here the spectral index of synchrotron emission is assumed to be $\alpha=-1$ 
for simplicity, but the discussions below will not be affected if the spectral 
index is different.  

\begin{figure*}
\centering
\includegraphics[bb=40 266 600 494,clip,width=0.94\textwidth]{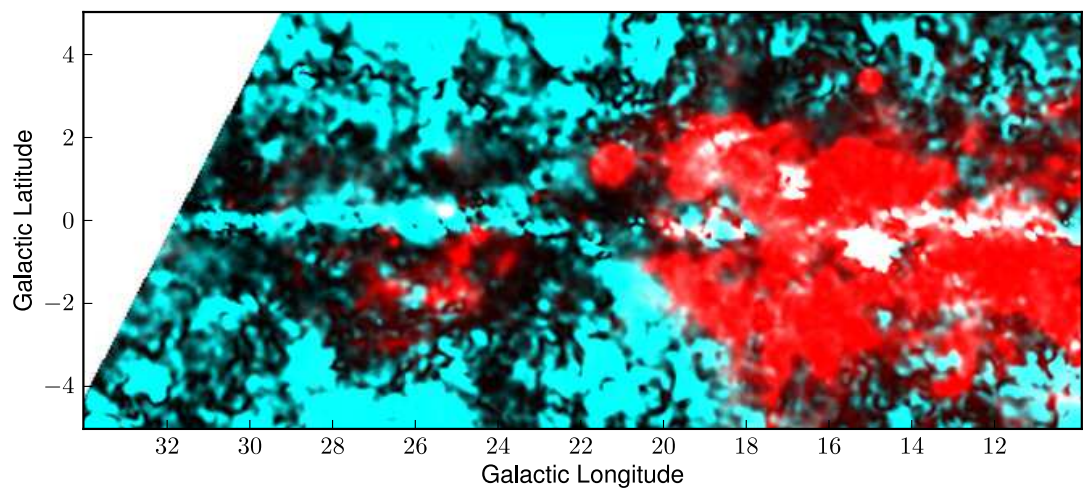}
\caption{Polarised intensity from the Parkes 2.3~GHz data in cyan overlaid with 
H$\alpha$ intensity from \citet{fin03} in red.}
\label{pihii}
\end{figure*}

According to Equation (\ref{eq:ipi}), we shall keep $I$ constant and let $p$ 
vary in order to generate polarised patches which do not correspond to smooth 
total intensity. There are three ways to achieve this: 
\begin{itemize}
\item Regions where regular magnetic fields dominate produce polarised patches, 
      while regions where isotropic random magnetic fields ($\sigma_x=\sigma_y$)
      dominate cause low polarisation voids or canals. For all the regions, the 
      sum of squared regular and random fields are similar so that the total 
      intensity is smooth. Since the polarised patches correspond to regular 
      magnetic fields which are coherent over a large spatial scale and mainly 
      parallel to the plane \citep[e.g.][]{srwe08}, it is expected that 
      polarisation angles towards all the patches should be around zero at 
      4.8~GHz. This, however, is not supported by the polarisation angle map 
      presented by \citet{srh+11}. This scenario is thus not favoured.
\item The random magnetic field is dominant and anisotropic. If the field is 
      isotropic, no polarisation is generated. If towards some regions the 
      field is anisotropic, namely $\sigma_x\neq\sigma_y$, polarised patches 
      will be produced. The anisotropy can be caused by compression or shear 
      motions of gas \citep[][]{jlb+10,fbs+11,jf12}. 
\item The random magnetic field is dominant and isotropic. The field also has a 
      non-zero, such as Kolmogorov-like (spectral index is $-11/3$), energy 
      spectrum so that coherent structures such as polarised patches could be 
      formed over the inertial scale of the turbulence~\citep{sr09}. Canals 
      actually mark the overlap regions of patches where polarised emission 
      partly or completely cancels.
\end{itemize}
The last two scenarios are both possible and they both require the random 
magnetic fields to be much stronger than the regular fields.

\subsubsection{Polarised structures in the Parkes 2.3~GHz data}\label{dpparkes}

There is much less polarised emission near the Galactic plane in the Parkes 
2.3~GHz data than in the Urumqi 4.8~GHz data (Fig.~\ref{pi}), indicating strong 
depolarisation at 2.3~GHz. The depolarisation depends on frequency and 
therefore must stem from Faraday rotation effects \citep{sbs+98}. Faraday 
rotation is related to thermal electrons that can be traced by H$\alpha$ 
emission. We extract the H$\alpha$ image from the synthetic all-sky map by 
\citet{fin03} which is mainly from SHASSA~\citep{gmrv01} in this direction. In 
Fig.~\ref{pihii} we show the Parkes 2.3~GHz data overlaid with the H$\alpha$ 
emission. 

The anti-correlation between H$\alpha$ emission and polarised intensity is very 
pronounced in Fig.~\ref{pihii}. This indicates that the warm ionised gas traced 
by the H$\alpha$ emission acts as Faraday screens to impact the depolarisation, 
although strong extinction in the Galactic plane means only nearby ionised gas 
has been observed. Several H{\scriptsize II} regions bright at both radio total 
intensity (Fig.~\ref{toti}) and H$\alpha$ are located at distances of 2--3~kpc 
such as Sh 2-45 ($15\fdg0$, $-0\fdg7$) at about 2.2~kpc, Sh 2-49 
($16\fdg9$, $+0\fdg8$) at about 2.6~kpc and Sh 2-54 ($18\fdg6$, $+1\fdg9$) at 
about 2.5~kpc. However, the H{\scriptsize II} region at ($18\fdg2$, $-0\fdg3$) 
at a distance of 3.8~kpc is only visible in radio total intensity maps but not 
in H$\alpha$ probably due to extinction. All the distances are from the 
compilation by \citet{hhs09} with a Galactocentric solar radius of 
$R_0=8.5$~kpc. In the following discussions, we assume all the H$\alpha$ 
emission is from a distance of 2--3~kpc, which could be located in either the 
Sagittarius arm or the Scutum arm (Fig.~\ref{spiral}). A similar analysis has 
been carried out by \citet{ccs+13}.

\begin{figure*}
\centering
\includegraphics[angle=-90,width=0.9\textwidth]{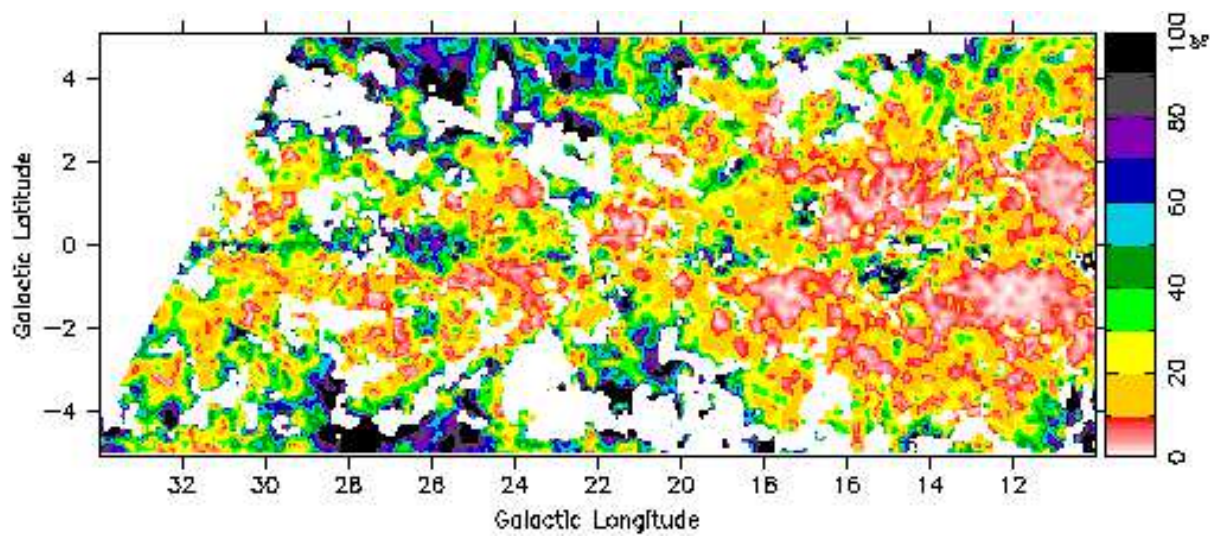}
\caption{Depolarisation factor (DF) calculated from the Parkes 2.3~GHz and 
the Urumqi 4.8~GHz data according to Equation~(\ref{eq:df}). Regions where the 
polarised intensity at 4.8~GHz is less than $5\sigma$ are masked.}
\label{dp}
\end{figure*}

Faraday rotation mainly causes two types of depolarisation relevant in our 
observations: depth depolarisation and beam depolarisation \citep{sbs+98}. 
Depth depolarisation happens where a synchrotron-emitting medium and a thermal 
medium are mixed. Emission from different depths experiences different Faraday 
rotation and thus has different polarisation angles. Adding up all the emission 
reduces polarisation. Beam depolarisation occurs where the thermal medium is in 
front of the emitting medium. If RMs of the thermal medium are uniform as a 
function of position in the sky, polarisation from the background is only 
rotated and there will be no depolarisation. However, if RMs exhibit 
irregularities over scales smaller than the observation beam width, averaging 
polarised intensity within the beam partly or completely cancels the 
polarisation. 
 
The H$\alpha$ emission in Fig.~\ref{pihii} mainly consists of discrete 
structures which can be related to H{\scriptsize II} regions. Therefore there 
is no mixture of thermal and non-thermal medium, and only beam depolarisation 
accounts for the depolarisation in the Parkes 2.3~GHz data. We can then 
conclude that the fluctuation scale for RMs is smaller than 6--9~pc given 
the beam width of $10\farcm75$ and the distance of 2--3~kpc. This fluctuation 
is probably induced by stellar sources in the arms with an energy injection 
scale of about 10~pc \citep{hbgm08}.  
 
To quantitatively describe depolarisation at 2.3~GHz, we define the 
depolarisation factor (DF) as
\begin{equation}\label{eq:df}
{\rm DF} = \frac{p_{2.3}}{p_{4.8}}\times\left(\frac{4.8}{2.3}\right)^\beta,
\end{equation}
where $p_{2.3}$ and $p_{4.8}$ are polarised intensity at 2.3~GHz and 4.8~GHz, 
respectively, and $\beta$ is spectral index for brightness temperature 
($T\propto\nu^\beta$). The depolarisation factor describes how much of the 
polarised emission extrapolated from 4.8~GHz is observed at 2.3~GHz. The 
smaller the depolarisation factor, the more significant the depolarisation is. 

\begin{figure*}
\centering
\includegraphics[angle=-90,width=0.45\textwidth]{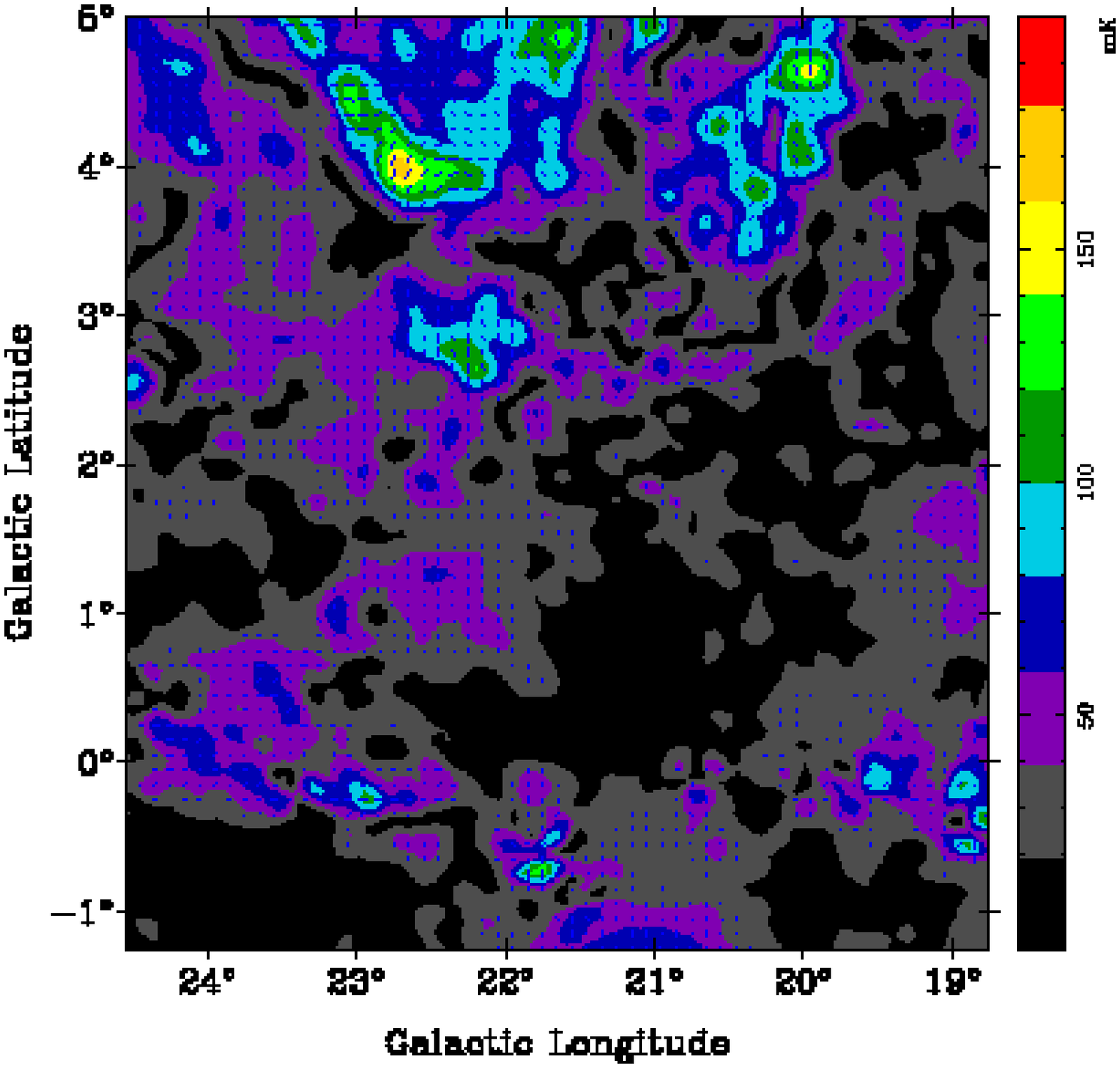}
\includegraphics[angle=-90,width=0.45\textwidth]{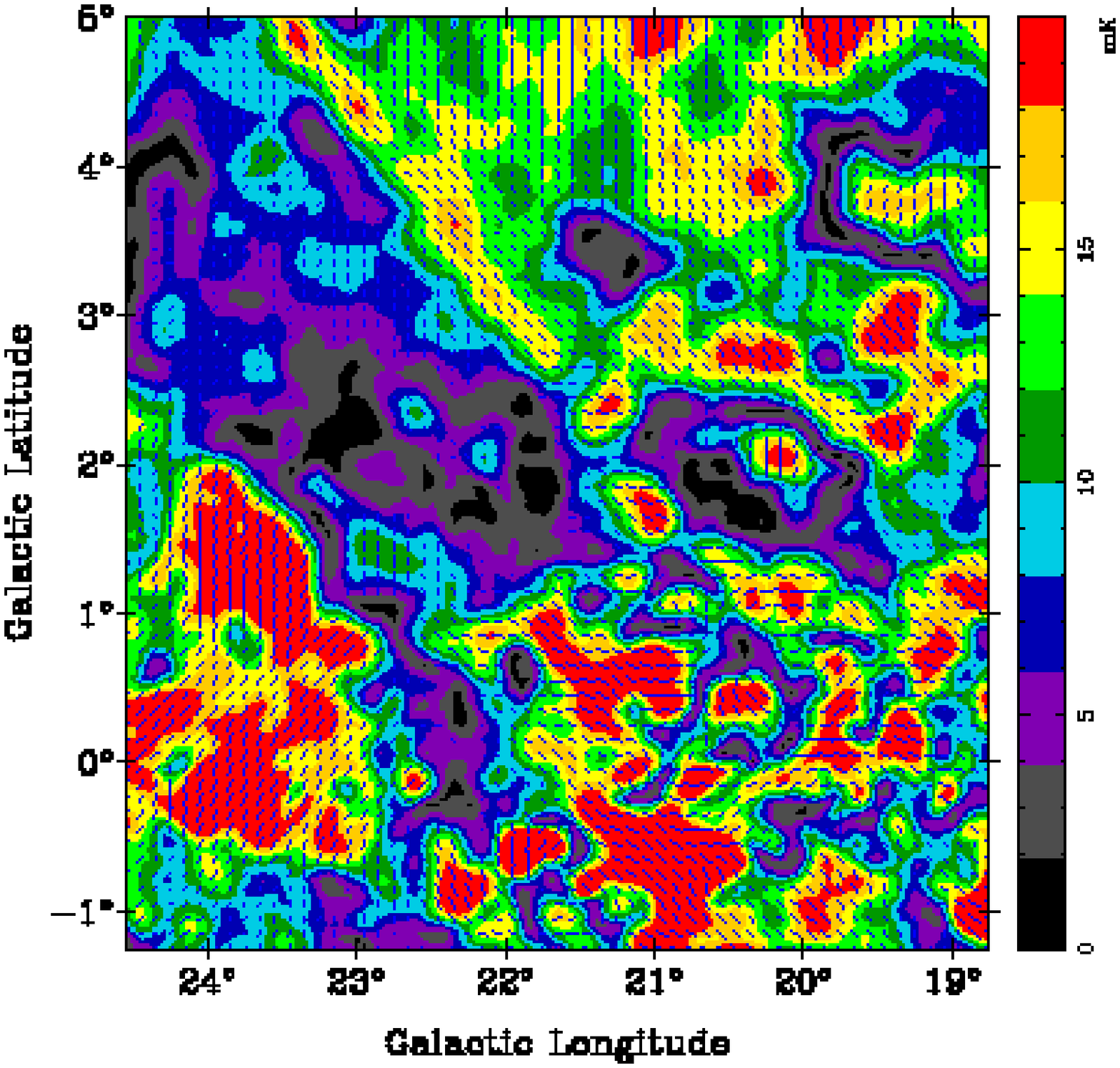}
\caption{Polarisation images covering the NPS low latitude extension from the 
Parkes 2.3~GHz data (left) and the Urumqi 4.8~GHz data (right) overlaid with 
bars indicating $B$ vectors (polarisation angle plus $90\degr$). The two colour 
ranges have been scaled so that polarised intensities corresponding to a 
spectral index $\beta=-3.1$ have the same colours.}
\label{nps6spass}
\end{figure*}

The DF maps calculated from the Parkes 2.3~GHz and the Urumqi 4.8~GHz data are 
shown in Fig.~\ref{dp}. Here we have followed \citet{srh+11} to take 
$\beta=-3.1$ which is calculated from the WMAP 22.8~GHz and 33~GHz polarised 
intensity. The mean depolarisation factor measured from the low polarisation 
area at 2.3~GHz overlapping with H$\alpha$ emission (Fig.~\ref{pihii}) is 
about 10\%. This means that 90\% of the polarised emission has been depolarised 
at 2.3~GHz and thus must be located behind the warm ionised gas at 2--3~kpc 
traced by H$\alpha$ emission (Fig.~\ref{pihii}). We also know that polarised 
emission is produced within 4~kpc from Sect.~\ref{piuru}. This means towards 
$10\degr<l<34\degr$ most of the polarised emission at 4.8~GHz comes from the 
Scutum arm at a distance of 3--4~kpc and the turbulent magnetic field dominates 
over the regular field in the arm to cause the patchy structures seen at 
4.8~GHz (Fig.~\ref{pi}).

For beam depolarisation, the fraction of polarisation is proportional to 
$\exp(-2\sigma_{\rm RM}^2\lambda^4)$ \citep[e.g.][]{bur66}, where 
$\sigma_{\rm RM}$ is the RM fluctuation within the beam. We can then represent 
the depolarisation factor as
\begin{equation}\label{eqdfrm}
{\rm DF} = \exp[-2\sigma_{\rm RM}^2(\lambda_{2.3}^4-\lambda_{4.8}^4)].
\end{equation}  
The depolarisation factor of about 10\% implies an RM dispersion  
$\sigma_{\rm RM}\approx65$~rad~m$^{-2}$ on scales of less than 6--9~pc. 
\citet{gdm+01} obtained a value of 35--50~rad~m$^{-2}$ towards the region 
($332\fdg5$, $+1\fdg2$) based on the depolarisation factor of 0.3--0.5 from 
the 2.4~GHz survey by \citet{dhjs97} and a fluctuation scale less than about 
2~pc based on the SGPS data. In our analysis, the depolarisation factor and 
thus the RM fluctuation are more robust as the high frequency data at 4.8~GHz 
are included. However, the fluctuation scale is loosely constrained to an upper 
limit at best, because of the much coarser angular resolution compared to the 
SGPS.

\subsubsection{Summary}

According to the discussion in Sects.~\ref{piuru} and \ref{dpparkes}, at 
4.8~GHz the observed polarised structures are intrinsic, and caused by random 
magnetic field in the emitting medium at a distance of 3--4~kpc. However, at 
2.3~GHz the observed polarised structures are extrinsic, and are caused by 
foreground Faraday screens composed of warm ionised gas at a distance of 
2--3~kpc.

\subsection{The branch extending from the North Polar Spur: a feature far away?}

The North Polar Spur (NPS) is one of the most prominent structures in the radio 
sky, protruding from the Galactic plane at longitude about $30\degr$ and 
reaching a latitude as high as $+80\degr$ \citep[e.g.][]{ber71}. It has been 
widely accepted that the NPS is part of an old supernova remnant at a distance 
of about 100~pc based on optical star light polarisation and the relation 
between surface brightness and distance for supernova remnants 
\citep[][and references therein]{sal83}. Recently, \citet{wol07} modelled the 
NPS as two shells at distances of 78~pc and 95~pc respectively. In contrast, 
\citet{sof00} interpreted the NPS as a shock front formed by about 10$^5$ 
supernovae 15 million years ago in the Galactic centre. \citet{bc03} proposed 
that the bipolar wind from the Galactic centre produces the NPS.

The key to understanding the nature of the NPS is to constrain its distance, 
for which polarisation observations can help. \citet{sr79} mapped the NPS in 
total intensity and presented the results in an image covering 
$16\degr<l<35\degr$ and $-6\degr<b<+20\degr$ at 1420~MHz with a resolution of 
$10\arcmin$. By applying unsharp masking, they found that a ridge extends from 
the NPS towards lower Galactic latitudes and splits into two branches at 
$(22\fdg5,\,+3\fdg6)$ and one branch reaches $(21\degr,\,+2\degr)$. These two 
branches can be clearly seen in the Urumqi 4.8~GHz total intensity map but not 
well seen in the Parkes 2.3~GHz total intensity image because of confusion from 
strong background emission (Fig.~\ref{toti}). In Fig.~\ref{nps6spass} we 
display polarisation structure at 2.3 and 4.8 GHz in small areas including the 
total-intensity branches (which are indicated by arrows in Fig.~\ref{pi}). Note 
that polarised intensities scaled with a spectral index $\beta=-3.1$ have the 
same colour in both images. The spectral index is consistent with that for the 
NPS between 408~MHz and 1420~MHz derived by \citet{rrt04}. In the Urumqi 
4.8~GHz data, polarised emission from one of the branches can be firmly 
identified from $b=+5\degr$ to $b=+2\fdg6$. Whether the branch can be traced 
further is very uncertain because many polarised fragments appear below 
$b=+2\fdg6$. In the Parkes 2.3~GHz data, the part above the latitude of about 
$+3\fdg8$ roughly matches the 4.8~GHz data, but the rest has significantly 
lower intensities than that extrapolated from the 4.8~GHz data with 
$\beta=-3.1$. 

The depolarisation at 2.3~GHz below a latitude of about $+3\fdg8$ towards the 
branch extending from the NPS indicates that the branch is far away. We first 
investigate the cause for depolarisation. At 4.8~GHz, the magnetic field 
vectors well align with the ridge (Fig.~\ref{nps6spass}) meaning the intrinsic 
RM is very small, which rules out depth depolarisation at 2.3~GHz. Beam 
depolarisation is therefore the main mechanism for depolarisation. However, 
there is no strong H$\alpha$ emission above the latitude of about $+2\fdg6$ 
towards the branch (Fig.~\ref{pihii}). The warm ionised gas producing large RM 
fluctuations is probably further away than that at lower latitudes, so that 
extinction prevents detection of strong H$\alpha$ emission. This implies that 
the branch is located at a distance larger than 2--3~kpc. 

\begin{figure*}
\centering
\includegraphics[angle=-90,width=0.45\textwidth]{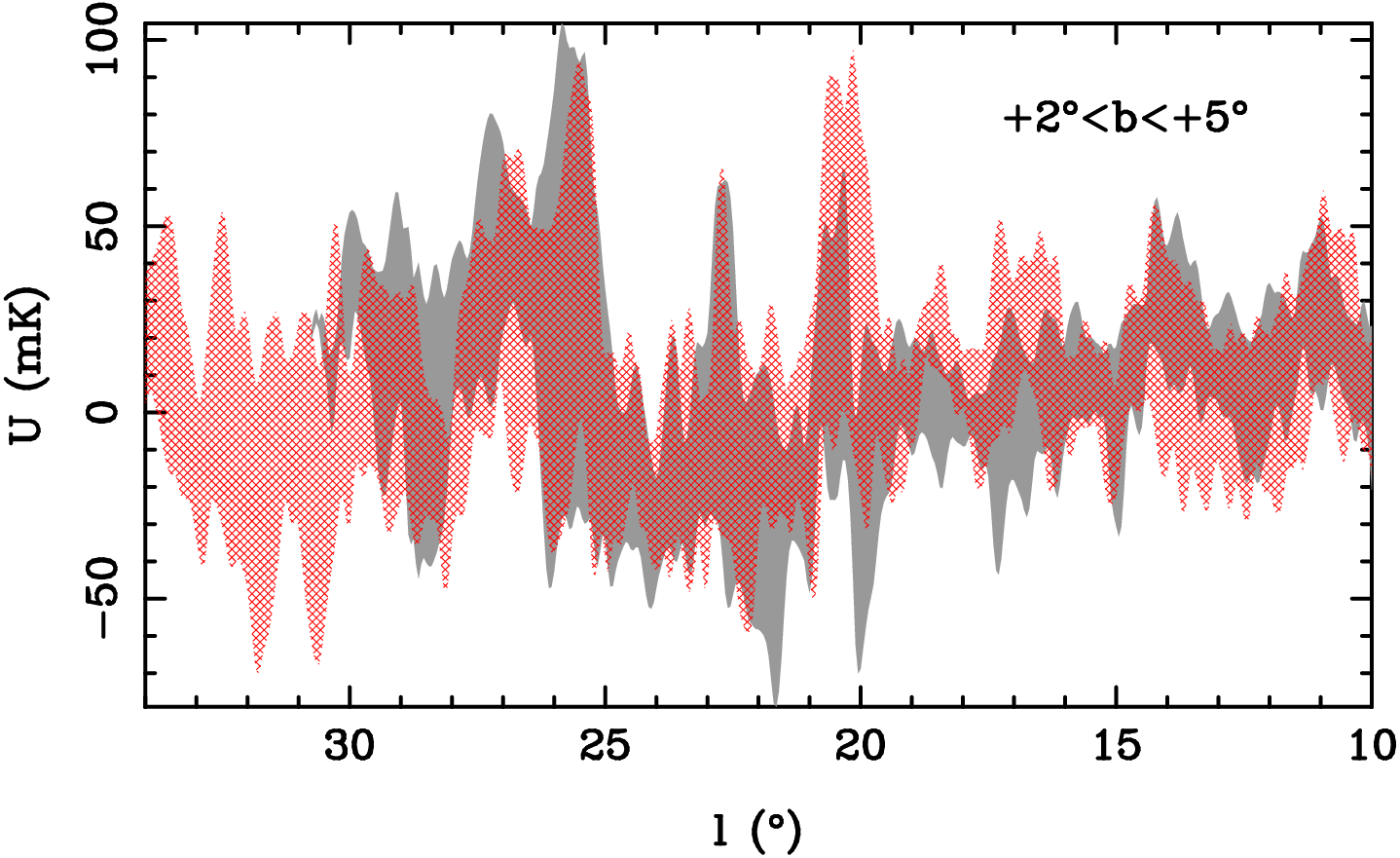}
\includegraphics[angle=-90,width=0.45\textwidth]{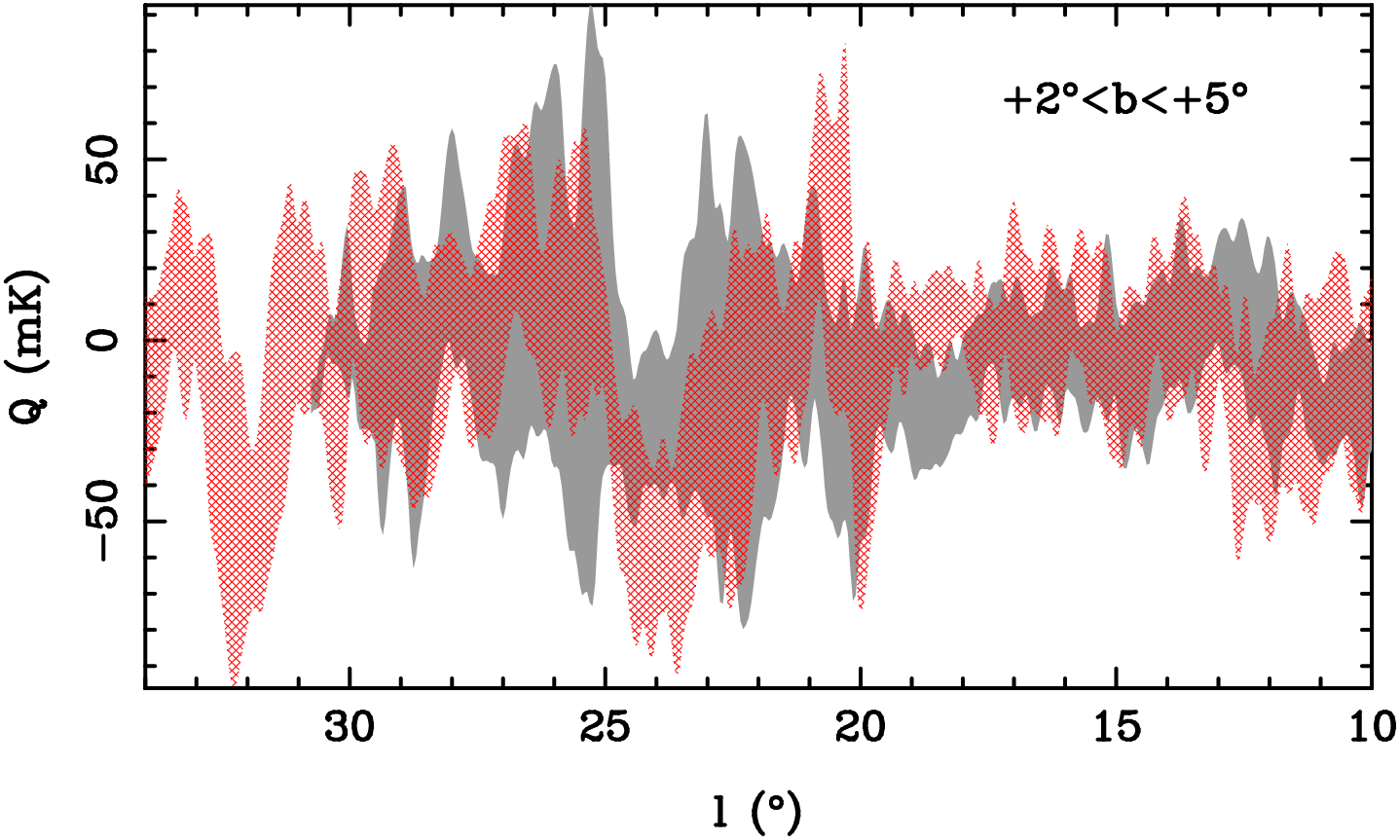}
\includegraphics[angle=-90,width=0.45\textwidth]{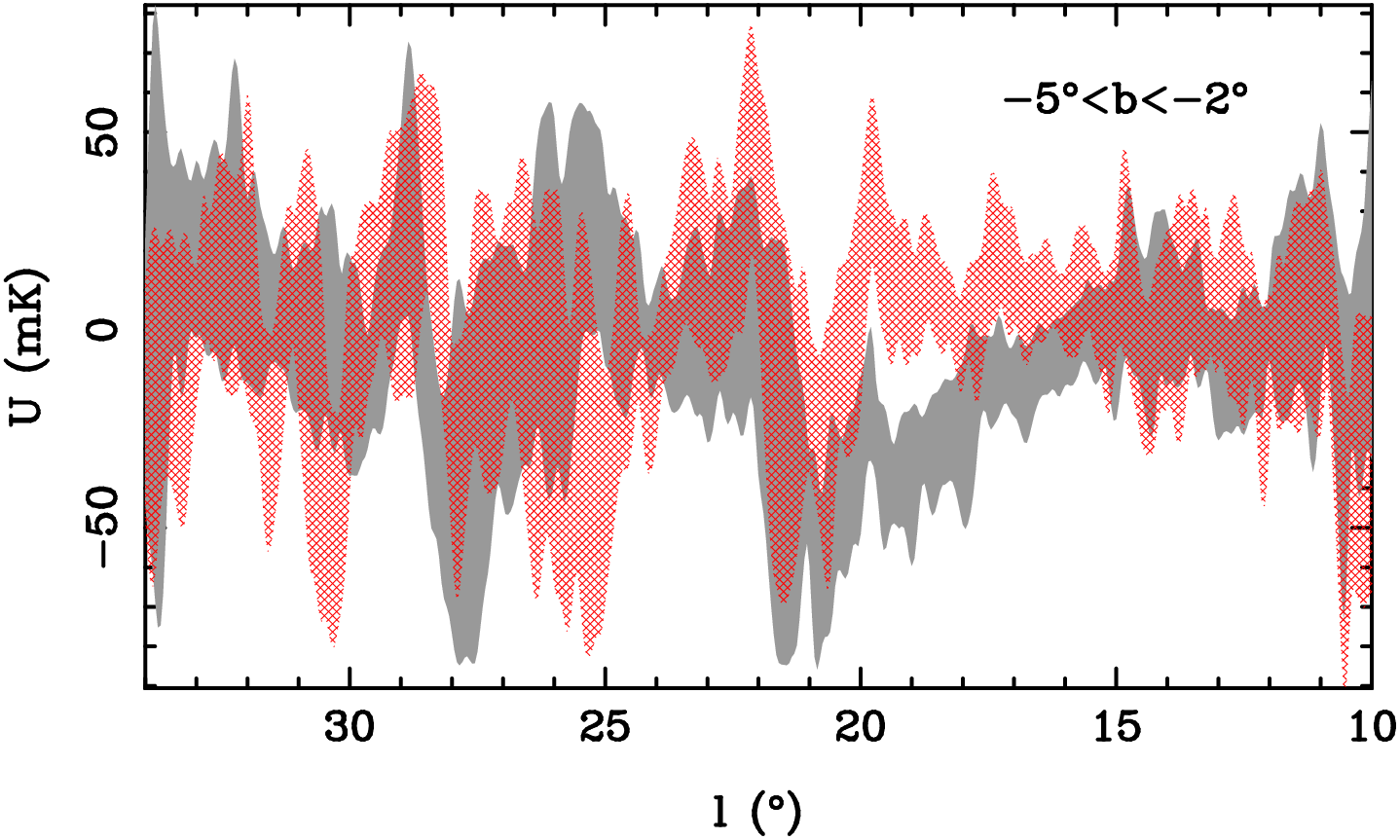}
\includegraphics[angle=-90,width=0.45\textwidth]{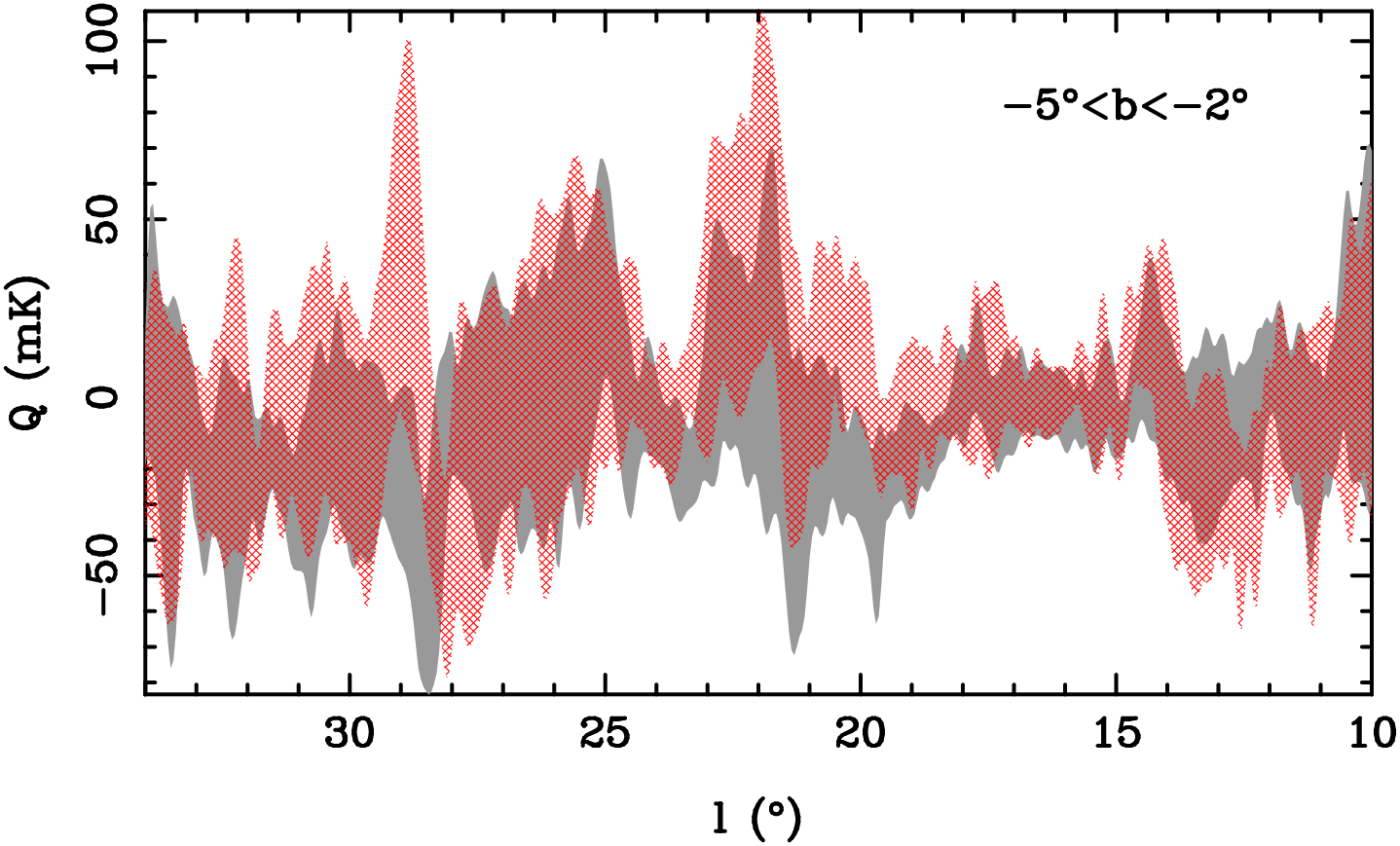}
\caption{Longitude profiles for $U$ and $Q$ taken from the Parkes 2.3~GHz data 
(gray) and the Effelsberg 2.7~GHz data (red).}
\label{comp}
\end{figure*}

It is not yet clear whether the branch extending from the NPS is physically 
associated with the NPS. If these two structures are physically connected, the 
distance to the NPS must be much further than the distance of 100~pc that was 
proposed decades ago. The NPS would intercept the Galactic plane at longitude 
of about $20\degr$ if we linearly extend the polarised branch at the Urumqi 
data towards the Galactic plane (Fig.~\ref{pi}). These facts agree very well 
with the scenario proposed by \citet{sof00}, namely that the NPS traces a shock 
front formed in the Galactic centre about 15~Myr ago and the tangential 
direction of the gaseous ring in the plane formed by the blast wave is about 
$\pm20\degr$. The Galactic wind model proposed by \citet{bc03} is not favoured 
because the NPS would be thermal and aligned with the Galactic centre in their 
model. 

It is also possible that the NPS and the branch are independent structures and 
uncorrelated with each other. In this case we cannot constrain the distance to 
the NPS. A detailed study of the NPS using GMIMS \citep{wlh+10} data is 
underway.  

\subsection{Polarisation in the Effelsberg 2.7~GHz data}

The baselines of the Effelsberg 2.7~GHz $U$ and $Q$ data were set by 
subtracting zeroth- or first-order fits so that the averages of all the 
pixels are zero \citep{drrf99}. This might limit the use of the dataset in 
studying large-scale polarised structures for which absolute calibration is 
required to warrant a reliable interpretation \citep[e.g.][]{rei06}. The Parkes 
2.3~GHz data have an absolute level and therefore allow us to assess to what 
extent the large-scale structures were missed in the Effelsberg data. 

We obtained longitude profiles of $U$ and $Q$ averaged over $+2\degr<b<+5\degr$ 
and $-5\degr<b<-2\degr$ for both the Parkes 2.3~GHz and Effelsberg 2.7~GHz data 
and the results are shown in Fig.~\ref{comp}. We have scaled the Effelsberg 
2.7~GHz data by using a spectral index $\beta=-3.1$. The Galactic plane was 
skipped due to strong depolarisation. Except for the discrepancy that appears 
in $U$ at $15\degr<l<20\degr$ and $-5\degr<b<-2\degr$ where a polarised 
filament was observed in the Parkes 2.3~GHz data, the profiles generally agree 
with each other. This is actually what we expect because the strong 
depolarisation near the plane breaks up large-scale coherent polarised 
structures as has been seen at lower frequencies \citep{hkd03b}. It is thus 
possible that towards the inner Galaxy the Effelsberg 2.7~GHz data without 
absolute calibration roughly approximate the Parkes 2.3~GHz data with absolute 
calibration. 

\subsection{Statistical studies}

\subsubsection{Structure function}
The second-order structure function for RMs of extragalactic radio sources 
has been used to study the turbulent properties of Galactic magnetic fields 
\citep[e.g.][]{scs84,ms96}. Below we show that the structure function for 
polarised intensity \citep[e.g.][]{hkd03a,xhr+11} can be used to diagnose 
whether the diffuse polarised emission is intrinsic or caused by Faraday 
screens. 

\begin{figure}
\centering
\includegraphics[angle=-90,width=0.46\textwidth]{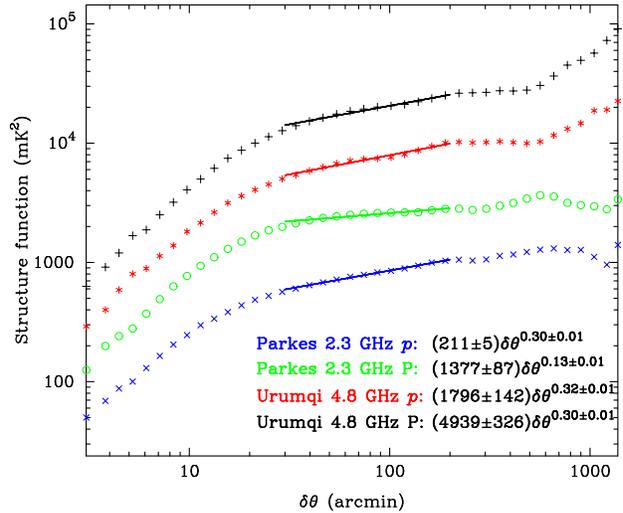}
\caption{Structure functions for complex polarised intensity $P=Q+iU$ and 
polarised intensity $p=\sqrt{U^2+Q^2}$ for the Urumqi 4.8~GHz and Parkes 
2.3~GHz data. The structure functions at 4.8~GHz have been scaled to 2.3~GHz 
with a spectral index $\beta=-3.1$.}
\label{sf}
\end{figure}

The structure functions for complex polarised intensity $P=Q+iU=p\exp(2i\psi)$ 
and polarised intensity $p=|P|$ are defined as
\begin{equation}\label{sfequ}
\begin{array}{r@{}c@{}l}
{\rm SF}_P(\delta\theta) &\,\,\equiv\,\,& \langle|P(\mathbf{\theta})-P(\mathbf{\theta+\delta\theta})|^2\rangle\\[3mm]
{\rm SF}_p(\delta\theta) &\,\,\equiv\,\,& \langle\left[p(\mathbf{\theta})-p(\mathbf{\theta+\delta\theta})\right]^2\rangle,
\end{array}
\end{equation}
where $\delta\theta$ is the angular separation between two lines of sight and 
$\langle\ldots\rangle$ stands for ensemble average. The formulae above can be 
further expanded to
\begin{equation}\label{sfequ_expand}
\begin{array}{r@{}c@{}l}
{\rm SF}_P(\delta\theta) &\,\,=\,\,& 2\langle p^2\rangle\\[3mm]
&\multicolumn{2}{l}{\,\,-2\langle p(\mathbf{\theta})p(\mathbf{\theta+\delta\theta})\cos\left\{2\left[\psi(\mathbf{\theta})-\psi(\mathbf{\theta+\delta\theta})\right]\right\}\rangle}\\[3mm]
{\rm SF}_p(\delta\theta) &\,\,=\,\,& 2\langle p^2\rangle-2\langle p(\mathbf{\theta})p(\mathbf{\theta+\delta\theta})\rangle,
\end{array}
\end{equation}
where $\langle p^2\rangle=\langle p^2(\mathbf{\theta})\rangle=\langle p^2(\mathbf{\theta+\delta\theta})\rangle$. 

We calculated ${\rm SF}_P$ and ${\rm SF}_p$ from the Parkes 2.3~GHz and Urumqi 
4.8~GHz data according to Equation~(\ref{sfequ}) and show the results in 
Fig.~\ref{sf}. The regions influenced by instrumental polarisation have been 
discarded for the calculation. We also scaled the structure functions at 4.8~GHz 
by $[(2.3/4.8)^\beta]^2$ with $\beta=-3.1$ to compare with those at 2.3~GHz. At 
angular separations less than about $15\arcmin$, the structure functions 
basically represent the smoothing effect of the observation beam. At angular 
separations larger than about $5\degr$, the structure functions cannot be well 
constrained. We therefore focus on the structure functions over the scales in 
between. Over the angular range $30\arcmin<\delta\theta<200\arcmin$, we can 
make linear fits to all the structure functions in logarithmic scales, and the 
results are shown in Fig.~\ref{sf}. 

If the observed polarisation is intrinsic, there will be no correlation between 
polarised intensity and polarisation angle. We can rewrite the structure 
function of complex polarised intensity by separating the averages in the 
2nd term in Equation~(\ref{sfequ_expand}) as
\begin{equation}\label{sfeqP_expand}
\begin{array}{r@{}c@{}l}
{\rm SF}_P(\delta\theta) &\,\,=\,\,& 2\langle p^2\rangle\\[3mm]
&\multicolumn{2}{l}{-2\langle p(\mathbf{\theta})p(\mathbf{\theta+\delta\theta})\rangle\langle\cos\left\{2\left[\psi(\mathbf{\theta})-\psi(\mathbf{\theta+\delta\theta})\right]\right\}\rangle}\\[3mm]
&\,\,=\,\,& 2\langle p^2\rangle - 2\langle p(\mathbf{\theta})p(\mathbf{\theta+\delta\theta})\rangle\times\\[3mm]
&\multicolumn{2}{l}{\quad\left(\langle 1-2\sin^2\left[\psi(\mathbf{\theta})-\psi(\mathbf{\theta+\delta\theta})\right]\rangle\right)}\\[3mm]
&\,\,\approx\,\,& 2\langle p^2\rangle - 2\langle p(\mathbf{\theta})p(\mathbf{\theta+\delta\theta})\rangle\times\\[3mm]
&\multicolumn{2}{l}{\quad\left(1-2\langle\left[\psi(\mathbf{\theta})-\psi(\mathbf{\theta+\delta\theta})\right]^2\rangle\right)}\\[3mm]
&\,\,=\,\,& {\rm SF}_p(\delta\theta) + 4\langle p(\mathbf{\theta})p(\mathbf{\theta+\delta\theta})\rangle\times{\rm SF}_\psi(\delta\theta).
\end{array}
\end{equation}
Here ${\rm SF}_\psi(\delta\theta)$ is the structure function for polarisation 
angles and it is assumed that $|\psi(\mathbf{\theta})-\psi(\mathbf{\theta+\delta\theta})|$ is small. The assumption is reasonable because we only focus on 
coherent structures over angular scales of $30\arcmin<\delta\theta<200\arcmin$.
We then substitute $\langle p(\mathbf{\theta})p(\mathbf{\theta+\delta\theta})\rangle$ according to Equation~(\ref{sfequ}) and derive the following
\begin{equation}\label{sfP}
{\rm SF}_P(\delta\theta)= {\rm SF}_p(\delta\theta) + \left(4\langle p^2\rangle-2\,{\rm SF}_p(\delta\theta)\right)\times{\rm SF}_\psi(\delta\theta).
\end{equation}
Generally it is expected that ${\rm SF}_p$ and ${\rm SF}_\psi$ should have the 
same slopes and it is obvious that $2\langle p^2\rangle>{\rm SF}_p$, therefore 
we conclude that ${\rm SF}_P$ should have a similar slope as ${\rm SF}_p$ but 
a much larger amplitude than ${\rm SF}_p$.

On the contrary, if the observed polarisation structures are caused by Faraday 
screens with beam depolarisation, polarised intensity will be anti-correlated 
with variation of polarisation angles as they are both related to RM 
fluctuations of the Faraday screens. In this case, the average term in 
${\rm SF}_P$ in Equation~(\ref{sfequ}) cannot be separated. We shall expect 
larger amplitude but flatter slope for ${\rm SF}_P$ than for ${\rm SF}_p$ as 
most of the intrinsic structures are smeared out by Faraday screens.  
   
From Fig.~\ref{sf} we can see at 4.8~GHz ${\rm SF}_P$ and ${\rm SF}_p$ have 
similar slopes, while at 2.3~GHz ${\rm SF}_P$ has much shallower slope than 
${\rm SF}_p$. This means that polarisation from the Urumqi 4.8~GHz data is 
intrinsic and polarisation from the Parkes 2.3~GHz data is caused by Faraday 
screens, which supports the discussion in Sect.~\ref{polgen}. We also note that 
${\rm SF}_p$ at both frequencies have similar slopes. This is because polarised 
intensities at these two frequencies differ only by the depolarisation factor. 

\begin{figure*}
\centering
\includegraphics[angle=-90,width=0.9\textwidth]{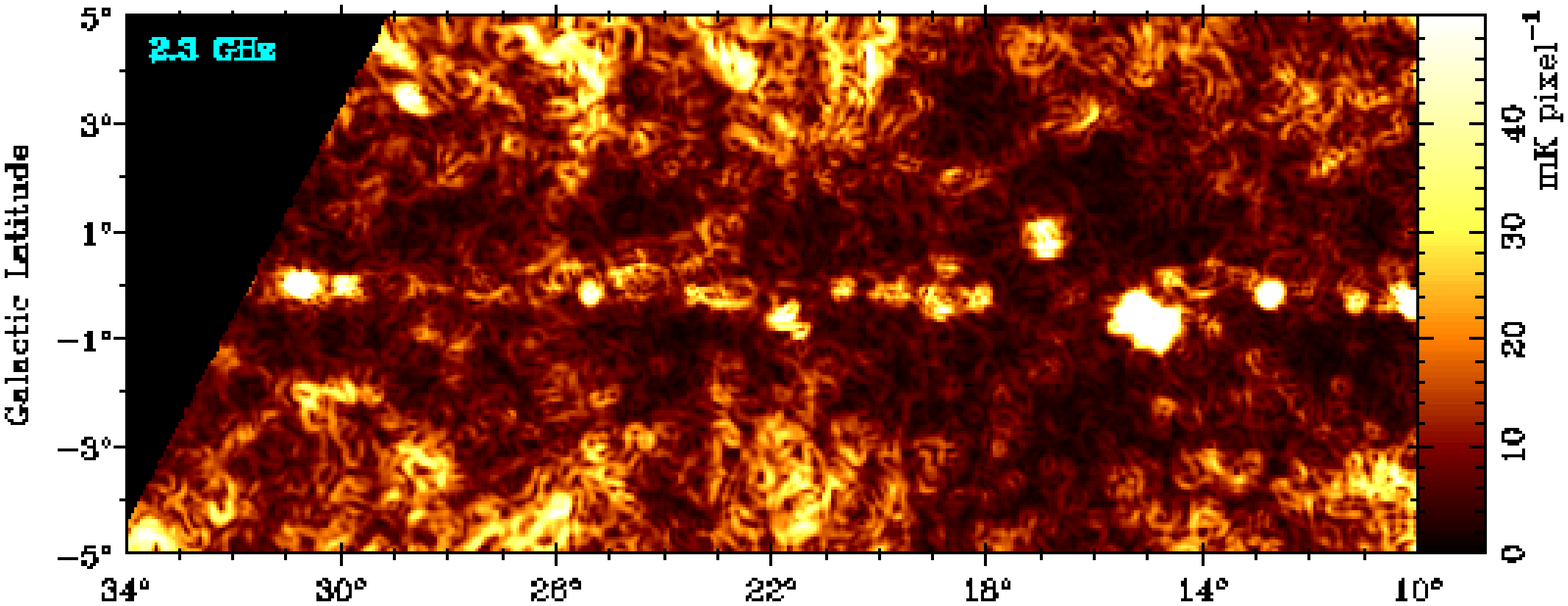}\\[2mm]
\includegraphics[angle=-90,width=0.9\textwidth]{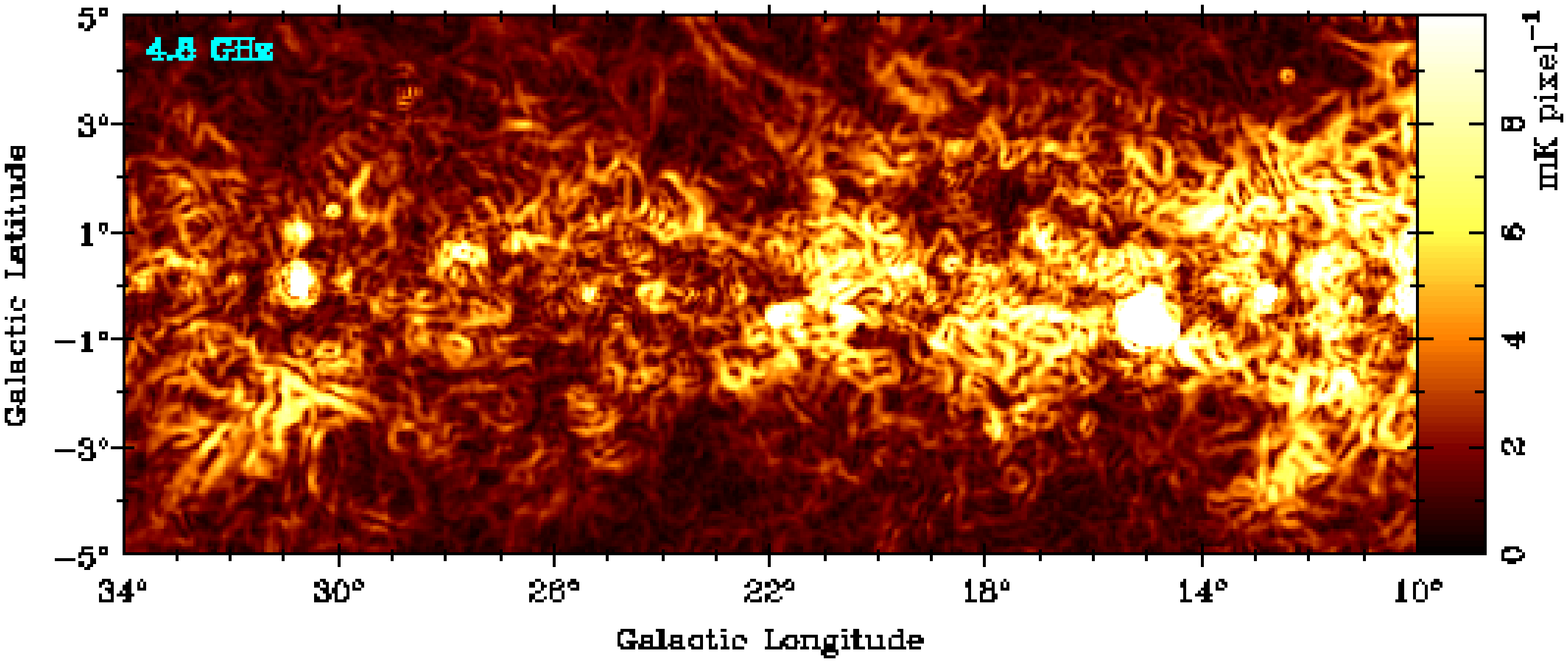}
\caption{Gradient maps ($|\nabla P|$) calculated from the Parkes 2.3~GHz data 
(top panel) and the Urumqi 4.8~GHz data (bottom panel) according to 
Equation~(\ref{eq:grad}).}
\label{gradmap}
\end{figure*}

\subsubsection{Gradient}

The spatial gradient of complex polarised intensity was first proposed and 
applied to SGPS data by \citet{ghb+11} revealing many filamentary structures 
corresponding to cusps or jumps in the turbulent interstellar medium. Following 
\citet{ghb+11}, we calculated the gradient maps for the Parkes 2.3~GHz and 
Urumqi 4.8~GHz data as    
\begin{equation}\label{eq:grad}
|\nabla P|=\sqrt{\left(\frac{\partial U}{\partial x}\right)^2+
\left(\frac{\partial U}{\partial y}\right)^2
+\left(\frac{\partial Q}{\partial x}\right)^2
+\left(\frac{\partial Q}{\partial y}\right)^2
}.
\end{equation}
The results are shown in Fig.~\ref{gradmap}.
 
In the gradient maps, all the bright structures except for discrete sources 
near $|b|=0\degr$ caused by instrumental polarisation are at latitudes 
$|b|>2\degr$ for the Parkes 2.3~GHz data. On the contrary, the prominent 
structures are mainly close to the Galactic plane for the Urumqi 4.8~GHz data. 
The structures in the gradient maps basically follow the distribution of 
polarised intensity (Fig.~\ref{pi}). Unlike the gradient map from the SGPS 
survey \citep{ghb+11}, the contrast between the bright and weak features is 
much larger for the gradient maps from the Parkes 2.3~GHz and Urumqi 4.8~GHz 
data.   

\begin{figure}
\includegraphics[angle=-90,width=0.48\textwidth]{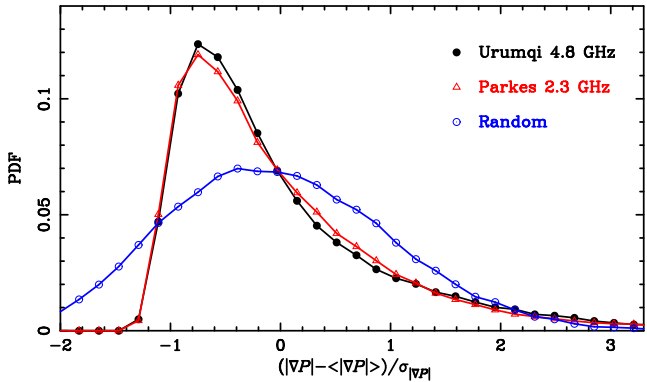}
\caption{Probability distribution functions for normalised $|\nabla P|$ 
calculated from the Urumqi 4.8~GHz data, the Parkes 2.3~GHz data, and a 
simulation containing only random noise.}
\label{grad}
\end{figure}

\citet{blg12} has proposed many statistical methods to infer turbulence 
properties based on the gradient maps. One of the probes is the probability 
distribution function (PDF) which seems to be a good indicator of sonic Mach 
number of the interstellar turbulence. Following \citet{blg12} we first shifted 
and normalised the gradients from both the Urumqi 4.8~GHz and Parkes 2.3~GHz 
data as $(|\nabla P|-<|\nabla P|>)/\sigma_{|\nabla P|}$ and then derived their 
PDFs. The results are shown in Fig.~\ref{grad}. For reference we also plot the 
result from simulated data with only random noise. The PDFs from these two 
observation data are very similar with a skewness of 1.7 and a kurtosis of 3.7 
for the Urumqi 4.8~GHz data and a skewness of 1.9 and a kurtosis of 5.6 for the 
Parkes 2.3~GHz data, and both are significantly skewed in comparison with the 
PDF from the random noise. 

Although the PDFs of the gradients from the Parkes 2.3~GHz and the Urumqi 
4.8~GHz data are similar, their physical meanings are totally different. The 
analysis by \citet{ghb+11} and \citet{blg12} has presumed that the observed 
polarised structures are caused by Faraday screens. For the data they 
considered, the gradients were able to trace the jumps and cusps in the 
foreground Faraday screens caused by MHD turbulence. If the polarised 
structures are intrinsic, the gradient analysis cannot be applied. The PDF of 
the gradients from the Parkes 2.3~GHz data indicate that the turbulence in the 
foreground Faraday screens are trans-sonic type by comparing Fig.~\ref{grad} 
with Fig.~6 of \citet{blg12} and referring to Fig.~7 of \citet{blg12}. However, 
it is not clear what the PDF of the gradients from the Urumqi 4.8~GHz data 
mean, as the polarised structures are intrinsic. 

\section{Conclusions}

By comparing polarised images covering $10\degr<l<34\degr$ and $|b|<5\degr$ 
at 2.3~GHz from S-PASS and at 4.8~GHz from the Sino-German $\lambda$6~cm 
polarisation survey, we obtain the following results:
\begin{itemize}
\item Polarised structures seen at 4.8~GHz are intrinsic, and are caused by 
      random magnetic fields. Most of the polarised emission originates from 
      the Scutum arm at a distance of 3--4~kpc.
\item Polarised structures at 2.3~GHz are caused by foreground Faraday screens 
      consisting of warm ionised gas at a distance of 2--3~kpc. These screens 
      have an RM fluctuation of about 65~rad~m$^{-2}$ over scales smaller than 
      6--9~pc, so that they cause nearly complete beam depolarisation.
\item The low-latitude branch extending from the North Polar Spur is at a 
      distance more than 2--3~kpc. If the branch and the NPS are physically 
      associated, the NPS cannot be local but likely associated with the 
      Galactic bulge, possibly tracing a shock front formed in the Galactic 
      centre. 
\item Structure functions of complex polarised intensity and polarised 
      intensity are capable of indicating whether observed polarised 
      structures are intrinsic or are caused by foreground Faraday screens. 
\item Gradients of the 2.3~GHz data indicate that turbulence in the warm 
      ionised medium is transonic. Gradients cannot be used to interpret the 
      4.8~GHz data in this way because the polarised structures seen at this 
      higher frequency are intrinsic.
\end{itemize}

\section*{Acknowledgements}

We thank Roland Kothes, Tom Landecker, Don Melrose and Wolfgang Reich for 
fruitful discussions. We also thank Tom Landecker and Roland Crocker for 
commenting on the draft. XHS, BMG and CRP were supported by the Australian 
Research Council through grant FL100100114. Parts of this research were 
conducted by the Australian Research Council Centre of Excellence for All-sky 
Astrophysics (CAASTRO), through project number CE110001020. This work is part 
of the research programme 639.042.915, which is (partly) financed by the 
Netherlands Organisation for Scientific Research (NWO). This work has been 
carried out in the framework of the S-band Polarisation All Sky Survey (S-PASS) 
collaboration. The Parkes Radio Telescope is part of the Australia Telescope 
National Facility, which is funded by the Commonwealth of Australia for 
operation as a National Facility managed by CSIRO.

\bibliographystyle{mn2e.longauthors}
\bibliography{bibtex.bib}
\end{document}